\documentclass[a4paper,12pt]{article}

\usepackage{indentfirst}

\usepackage{amsfonts,amsmath,amssymb,bm,a4wide,graphicx,cite}

\fussy                             

\begin{document}

\title{Spin light of neutrino in gravitational fields}

\author{
 M.Dvornikov
 \\
 e-mail: maxim\_dvornikov@aport.ru
 \\
 A.Grigoriev
 \\
 e-mail: alex.grigoriev@aport.ru
 \\
 A.Studenikin
 \\
 e-mail:
  studenik@srd.sinp.msu.ru
  \\
  \\
Department of Theoretical Physics,
  \\
Moscow State University,
  \\
  119992  Moscow, Russia}
  \date{}
  \maketitle
%


\begin{abstract}
We predict a new mechanism for the spin light of neutrino
($SL\nu$) that can be emitted by a neutrino moving in
gravitational fields. This effect is studied on the basis of the
quasiclassical equation for the neutrino spin evolution in a
gravitational field. It is shown that the gravitational field of a
rotating object, in the weak-field limit, can be considered as an
axial vector external field which induces the neutrino spin
procession. The corresponding probability of the neutrino spin
oscillations in the gravitational field has been derived for the
first time. The considered in this paper $SL\nu$ can be produced
in the neutrino spin-flip transitions in gravitational fields. It
is shown that the total power of this radiation is proportional to
the neutrino gamma factor to the fourth power, and the emitted
photon energy, for the case of an ultra relativistic neutrino,
could span up to gamma-rays. We investigate the $SL\nu$ caused by
both gravitational and electromagnetic fields, also accounting for
effects of arbitrary moving and polarized matter, in various
astrophysical environments. In particular, we discuss the $SL\nu$
emitted by a neutrino moving in the vicinity of a rotating neutron
star, black hole surrounded by dense matter, as well as by a
neutrino propagating in the relativistic jet from a quasar.
\end{abstract}

\section{Introduction}

It is well known that a neutrino with non-zero mass has
non-trivial electromagnetic properties \cite{FujShr80} (for the
recent study of the electromagnetic properties of a massive
neutrino see \cite{DvoStuPRD04,DvoStuJETP04}). In particular, the
Dirac massive neutrino possesses non-vanishing magnetic moment.
Even the massive Majorana neutrino can have transitional magnetic
moment.

In a series of our papers we have developed the Lorentz invariant
approach to neutrino spin (and also flavour
\cite{LikStu95,GriLobStuPLB02}) evolution in different
environments accounting for the presence of not only
electromagnetic fields \cite{EgoLobStuPLB00,DvoStuYF01_04} and
weakly interacting with neutrino matter \cite{LobStuPLB01} but
also for other types of neutrino non-derivative interactions
\cite{DvoStuJHEP02}. On this basis we have also predicted
\cite{LobStuPLB03} the possibility for a new mechanism of
electromagnetic radiation by neutrino moving in background matter
and/or electromagnetic fields. The new mechanism of
electromagnetic radiation ( we have named \cite{LobStuPLB03} this
radiation \emph{"spin light of neutrino"} ($SL\nu$) ) originates
from the neutrino spin precession that can be produced whether by
weak interactions with matter or by electromagnetic interactions
with external electromagnetic fields. The latter possibility was
also considered before in \cite{BorZhukTernSPJ88}. A review and
some new results on our studies of neutrino oscillations in matter
and external fields is given in {\cite{StuYF04}.

 As it is shown in \cite{LobStuPLB03}, the total power of the
$SL\nu$ is not washed out even when the emitted photon refractive
index in the background matter is equal to unit. That is why the
$SL\nu$ can not be considered as the neutrino Cherenkov radiation
(see, for example, \cite{IoaRaf97} and references therein). The
total power of the $SL\nu$ in matter is proportional to the fourth
power of the matter density and the neutrino Lorentz factor. The
$SL\nu$ is strongly beamed in the direction of neutrino
propagation and is confined within a small cone given by
$\vartheta\sim\gamma^{-1}$. The energy range of emitted $SL\nu$
photons, for the case of the relativistic neutrino, could span up
to gamma-rays. These properties of $SL\nu$ enables us to predict
that this radiation should be important in different astrophysical
environments (quasars, gamma-ray bursts etc) and in dense plasma
of the early Universe.

 In this paper we should like to
introduce a new type of the $SL\nu$ that could originate from the
neutrino spin precession induced by a gravitational field in
different astrophysical and cosmological settings. This study is
based on consideration of the neutrino spin evolution problem in
gravitational fields. We also consider the neutrino spin
oscillations in gravitational fields and derive for the first time
the corresponding probability.

The problem of the neutrino flavour oscillations and neutrino spin
evolution in the presence of gravitational fields have been
studied in many papers. The gravitational effect on flavour
neutrinos from collapsing stars was considered in
\cite{AhlBurGRG96}. Neutrino spin procession and oscillations in
gravitational fields were also studied in
\cite{PirRoyWudPRD96,CarFulPRD97,AthNiePRD00,CasMonPRD94,NiePalPRD58,CaiPapPRL91}.
The important obtained result of these studies is that if the
contribution of gravitational interaction is diagonal in spin
space (that is the case of gravitational fields of non-rotating
massive objects), then gravity cannot produce the neutrino spin
procession on its own. As it is pointed out in \cite{CarFulPRD97},
the off-diagonal terms in neutrino spin space could appear from
the interaction of a neutrino magnetic moment with a magnetic
field.

A rather detailed discussion on various aspects of the neutrino
spin and chiral dynamics in the presence of gravitational fields
can be found in the recent papers
\cite{SmiBlois96,Sin_gr_qc04,Muk_gr_qc04,CroGiuMorPRD04,Ahl_Kha_0405112}
where, in particular, the case of a strong field in a
Schwarzschild space-time background have been also considered. The
case of a weak field was also studied in \cite{CasMonPRD94}.
However, the possibility of the electromagnetic radiation due to
the neutrino spin procession in a gravitational field has never
been discussed before.

Here below we apply the Lorentz invariant approach, based on the
generalization of the quasiclassical Bargmann-Michel-Telegdi
equation for the case of different non-derivative interactions of
neutrino with external fields
\cite{EgoLobStuPLB00,LobStuPLB01,DvoStuJHEP02}, to the study of
the neutrino spin evolution in weak gravitational fields. We found
out that  weak gravitational fields of a rotating source enters
the neutrino spin evolution equation as an axial-vector field,
that, as it has been shown in \cite{DvoStuJHEP02}, produces the
spin precession. Then, using the main idea of \cite{LobStuPLB03}
that a neutrino with spin processing in the external environment
should radiate electromagnetic waves, we investigate this spin
light of neutrino in gravitational fields in the vicinity of the
rotating neutron star, the accretion disk of a rotating black hole
and in the relativistic jet of a quasar.

\section{Neutrino spin evolution equation in gravitational fields}

Following the study of ref.\cite{CasMonPRD94}, our starting point
is the Dirac equation for a neutrino in a curved space-time
\cite{DeW65BirrDav82}:
\begin{equation}
[i\gamma^{\mu}(x)D_{\mu}-m]\Psi=0,
\end{equation}
where $D_{\mu}=\partial_{\mu}+\Gamma_{\mu}$ is the covariant
derivative operator, $\Gamma_{\mu}=-(i/8)e^{\nu}_ae_{\nu
b;\mu}[\gamma^a,\gamma^b]$ is the spin connection, and $e^{\mu}_a$
are the tetrads. Latin indices refer to a local Minkowski frame,
while greek indices refer to general curvilinear coordinates. The
gamma matrices $\gamma^{{\mu}}=\gamma^ae^{{\mu}}_a$ satisfy the
conditions $\{\gamma^{\mu},\gamma^{\nu}\}=2g^{\mu\nu}$ and
$D_{\mu}\gamma^{\nu}=0$. The metric tensor $g_{\mu\nu}$ and the
tetrads $e^{\mu}_a$ are related by
\begin{equation}
  g_{\mu\nu}=\eta_{ab}e^a_{\mu}e^b_{\nu},
\end{equation}
where $\eta_{ab}$ is the Minkowski metric.

We consider below the limit of  a weak gravitational field which
is appropriate for the most of astrophysical systems. Indeed, we
are also restricted to the weak-field limit by our Lorentz
invariant approach in description of the neutrino spin evolution
on the basis of the generalized Bargmann-Michel-Telegdi equation.
Therefore, we perform our analysis within the linearized theory of
gravity by assuming that the metric can be written as
\begin{equation}
  g^{{\mu}{\nu}}=\eta^{{\mu}{\nu}}+2kh^{{\mu}{\nu}},
\end{equation}
and treat the second term in the right-hand side as a perturbation
over the flat Minkowski space. In this approach $h^{{\mu}{\nu}}$
is the gravitation field, and the quantity $k$ is related to the
Newton's constant $G_{N}$,
\begin{equation}
k= \sqrt{8\pi G_{N}}.
\end{equation}

As it has been shown in \cite{CasMonPRD94}, the corresponding
Hamiltonian for the considered linear approximation is
\begin{equation}\label{H1}
  \mathcal{H}=\gamma^0m(1-k\thinspace h^{00})+{\bm{\alpha}}{\mathbf{p}}-
  \frac{k}{2}\{h^{00},{\bm{\alpha}}{\bf p}{\}}_+ -
  \frac{k}{2}\{h^{ij},\alpha^ip^j{\}}_++k\thinspace\{{\bf h},{\bf p}{\}}_++
  k\left([\bf{\nabla}\times\mathbf{h}]
  \thinspace\mathbf{s}\right),
\end{equation}
where $\mathbf{p}=-i  \hbar\bm{\nabla}$,
$\bm{\alpha}=\gamma^0\bm{\gamma}$, and
$\mathbf{s}=\frac{1}{2}\bm{\Sigma}=\frac{1}{2}\gamma^0\gamma^5\bm{\gamma}$.
Note that in the limit of a weak gravitational field it is
possible to treat gravitational perturbations $h^{oi}=\bf h$ as an
external field contributing to the neutrino interaction
Lagrangian. Starting from the neutrino interaction Lagrangian
$\cal L$ accounting for general types of neutrino non-derivative
interactions with external fields,
\begin{align}
  -\mathcal{L}=&g_{s}s(x){\bar \nu}\nu+ g_{p}{\pi}(x){\bar
  \nu}\gamma^{5}\nu+ g_{v}V^{\mu}(x){\bar \nu}\gamma_{\mu}\nu+
  g_{a}A^{\mu}(x){\bar \nu}\gamma_{\mu}\gamma^{5}\nu
  \notag
  \\
  &+
  {{g_{t}}\over{2}}T^{\mu\nu}{\bar \nu}\sigma_{\mu\nu}\nu+
  {{g^{\prime}_{t}}\over{2}} \Pi^{\mu\nu}{\bar
  \nu}\sigma_{\mu\nu}\gamma_{5}\nu,
\end{align}
where $s, \pi, V^{\mu}=(V^{0}, \mathbf{V}), A^{\mu}=(A^{0},
\mathbf{A}), T_{\mu\nu}=(\mathbf{a}, \mathbf{b}),
\Pi_{\mu\nu}=(\mathbf{c}, \mathbf{d})$ are the scalar,
pseudoscalar, vector, axial-vector, tensor, pseudotensor fields,
respectively, and $g_{i}$ ($i=a$, $p$, $v$, $a$, $t$,
$t^{\prime}$) are the coupling constants,
$\sigma_{\mu\nu}=(i/2)(\gamma_{\mu}\gamma_{\nu}-
\gamma_{\nu}\gamma_{\mu})$, we get \cite{DvoStuJHEP02} the
following expression for the corresponding Hamiltonian:
\begin{align}
  \mathcal{H}=&g_{s}s\rho_{3}- ig_{p}\pi\rho_{2}+
  g_{v}(V^{0}-(\bm{\alpha}\mathbf{V}))- g_{a}(\rho_{1}A^{0}-(\bm{  \Sigma}\mathbf{A}))
  \notag
  \\
  &-
  g_{t}(\rho_{3}(\bm{\Sigma}\mathbf{b})+
  \rho_{2}(\bm{\Sigma}\mathbf{a}))- ig^{\prime}_{t}(\rho_{3}(\bm{  \Sigma}\mathbf{c})-
  \rho_{2}(\bm{\Sigma}\mathbf{d})),
  \label{H2}
\end{align}
where $\bm{\alpha}=\gamma^{0}\bm{\gamma}$, $\rho_1=-\gamma _{5}, \
\rho _{3}= \gamma ^{0},  \ \rho_{2}=i\rho_{1}\rho_{3}$.

The last term in equation (\ref{H1}),
\begin{equation}
  \mathcal{H}_\mathrm{spin}=
  k(\thinspace[\bm{\nabla}\times\mathbf{h}]\mathbf{s}),
  \label{Hspin}
\end{equation}
which is responsible for the neutrino spin interaction with a
gravitational field, can be identified  with an axial-vector part
of the general Hamiltonian given by (\ref{H2}) if we define the
axial-vector field $A^{\mu}=(0,\mathbf {A})$ by the following
relation
\begin{equation}
  g_a\mathbf{A}=\frac{k}{2}[\bm{\nabla}\times\mathbf{h}],
  \label{A}
\end{equation}
where $g_a$ is the coupling constant of this field to a neutrino.
Following the performed in \cite{DvoStuJHEP02} derivation of the
quasiclassical spin evolution equation accounting now only for the
axial-vector interaction given by (\ref{A}), we get the evolution
equation for the neutrino spin vector ${\bm {\zeta}}_{\nu}$  in
the considered gravitational field
\begin{equation}\label{eq}
  \frac{d\bm{\zeta}_\nu}{dt}=
  \frac{2}{\gamma}[\bm{\zeta}_\nu\times\mathbf{G}],
\end{equation}
where $t$ is time in the laboratory frame, and, as it follows from
eq.(8) of \cite{DvoStuJHEP02}, the vector $\bf G$ is
\begin{equation}\label{G}
  \mathbf{G}=-\frac{k}{2}
  \left\{
  [\bm{\nabla}\times\mathbf{h}]+
  \frac{\gamma}{1+1/\gamma}
  \bm{\beta}(\bm{\beta}[\bm{\nabla}\times\mathbf{h}])
  \right\}.
\end{equation}
Here $\bf \beta$ is the speed , $\gamma=E_{\nu}/m_{\nu}$ is the
Lorentz factor of the neutrino.

It should be noted that the derived neutrino spin evolution
equation (\ref{eq}) accounts only for the gravitational field,
however the possible effects of the neutrino interactions with the
background matter and electromagnetic fields are not included
here. The analogous neutrino spin evolution equation in the
presence of electromagnetic fields and matter in the Minkowski
space ( see refs. (\cite{EgoLobStuPLB00,DvoStuJHEP02,LobStuPLB03,
StuYF04}) is
\begin{equation}\label{S} {d\bm{\zeta}_{\nu}
\over dt}={2\mu \over \gamma} \Big[ {\bm{\zeta}_{\nu} \times
(\mathbf{B}_0+\mathbf{M}_0)} \Big],
\end{equation}
\begin{equation}\label{B_0}
\mathbf B_0=\gamma\Big(\mathbf B_{\perp} +{1 \over \gamma} \mathbf
B_{\parallel} + \sqrt{1-\gamma^{-2}} \Big[{\mathbf E_{\perp}
\times \mathbf n}\Big]\Big), \ \gamma = (1-\beta^2)^{-{1 \over
2}},
\end{equation}

\begin{equation}
  \mathbf {M_0}=\mathbf{M}_{0_{\parallel}}+\mathbf{M}_{0_{\perp}},
  \label{M_0}
\end{equation}
\begin{equation}
\begin{array}{c}
\displaystyle
  \mathbf{M}_{0_{\parallel}}=
  \gamma\bm{\beta}{n_0\over \sqrt {1- v_{e}^{2}}}
  \left\{
      \rho^{(1)}_{e}
      \left
          (1-{\mathbf {v}_e \bm{\beta} \over {1- {\gamma^{-2}}}}
      \right)
      \right.
-\\-
\displaystyle
      \rho^{(2)}_{e}\left.
      \left(
          \bm{\zeta}_{e}\bm{\beta} \sqrt{1-v^2_e}+
          {(\bm{\zeta}_{e}\mathbf{v}_e)(\bm{\beta}\mathbf{v}_e) \over
          1+\sqrt{1-v^2_e} }
      \right)
   {1 \over {1- {\gamma^{-2}}}}
   \right\},
\label{M_0_par}
\end{array}
\end{equation}
\begin{equation}\label{M_0_per}
\begin{array}{c}
\displaystyle
  \mathbf{M}_{0_{\perp}}=-\frac{n_{0}}{\sqrt {1-
  v_{e}^{2}}} \left\{ \mathbf{v}_{e_{\perp}}
     \left(
        \rho^{(1)}_{e}+\displaystyle\rho^{(2)}_{e}\frac
        {\bm{\zeta}_e\mathbf{v}_e} {1+\sqrt{1-v^2_e}}
     \right)
  +{\bm{\zeta}_{e_{\perp}}}\rho^{(2)}_{e}\sqrt{1-v^2_e} \right\},
\end{array}
\end{equation}
where   $\mathbf F_{\perp}$ and $\mathbf F_{\parallel}$ ($\mathbf
F= \mathbf B,\mathbf E$) are transversal and longitudinal (with
respect to the direction of neutrino motion $\mathbf{n}= {\bm
{\beta}}/ \beta$) electromagnetic field components in the
laboratory frame. For simplicity  we neglect here the neutrino
electric dipole moment, $\epsilon=0$, and also consider the case
when matter is composed of only one type of fermions (electrons).
The general case of $\epsilon\neq0$ and matter composed of
different types of leptons is discussed in our papers mentioned
above (\cite{EgoLobStuPLB00,DvoStuJHEP02,LobStuPLB03, StuYF04}).
Note that $n_0=n_{e}\sqrt {1-v^{2}_{e}}$ in (\ref{M_0_par}) and
(\ref{M_0_per}) is the invariant number density of matter given in
the reference frame for which the total speed of matter is zero.
The vectors $\mathbf v_e$, and ${\bm \zeta}_e \ (0\leqslant
|\mathbf \zeta_e |^2 \leqslant 1)$ denote, respectively, the speed
of the reference frame in which the mean momentum of matter
(electrons) is zero, and the mean value of the polarization vector
of the background electrons in the above mentioned reference
frame. The coefficients $\rho^{(1,2)}_e$ are calculated if the
neutrino interaction with matter is given. Therefore, within the
extended standard model supplied with $SU(2)$-singlet right-handed
neutrino $\nu_{R}$, we have
\begin{equation}\label{rho}
\rho^{(1)}_e={\tilde{G}_F \over {2\sqrt{2}\mu }}\,, \qquad
\rho^{(2)}_e =-{G_F \over {2\sqrt{2}\mu}}\,,
\end{equation}
where $\tilde{G}_{F}={G}_{F}(1+4\sin^2 \theta _W).$

From the neutrino evolution equation (\ref{S}) we have predicted
that the neutrino spin precession can be induced not only by
external electromagnetic fields but also by the neutrino weak
interaction with particles of the background matter. Now from the
neutrino spin evolution equation (\ref{eq}) in the gravitational
field it follows that the neutrino spin procession can appear if
the vector $\bm G$ is at least not zero. Thus, if the spin is
processing the off-diagonal metric components $h^{0i}$ have to
depend on the space coordinates. The appropriate example is
provided by the Kerr geometry that corresponds to the
gravitational field of a rotating with angular momentum $\bm L$
object. In this case
\begin{equation}
  \mathbf{h}=
  \frac{k}{8\pi}[\mathbf{L}\times\mathbf{r}],
\end{equation}
and the axial-vector field $\bm A$ can be written as
\begin{equation}\label{AKerr}
  g_a\mathbf{A}=
  G_N\frac{3\mathbf{r}(\mathbf{L}\mathbf{r})-r^2\mathbf{L}}{r^5}.
\end{equation}

Thus, we conclude that, as it follows from the neutrino spin
evolution equation (\ref{eq}), the spin procession can appear if
 the neutrino is moving in  a weak gravitational field of a rotating
object.

\section{Probability of neutrino spin oscillations in
gravitational fields}

Using the results of the previous section, it is possible to get
the corresponding neutrino spin oscillations Hamiltonian which can
be used in derivation of the neutrino spin oscillations
probability in the presence of the gravitational field.

Let us consider the case when the neutrino is moving along the
radial direction and decompose the vector $\mathbf{G}$ following
\begin{equation}\label{G}
\bf{G}= \bf{G}_{\parallel} + \bf{G}_{\perp},
\end{equation}
where
\begin{equation}\label{Gpar}
  \mathbf{G}_{\parallel}=-\gamma\frac{G_NL}{r^3}\mathbf{n}\cos{\theta},
\end{equation}
and
\begin{equation}\label{Gper}
  \mathbf{G}_{\perp}=\frac{1}{2}\frac{G_NL}{r^3}\mathbf{n}_{\perp}\sin{\theta}
\end{equation}
are the longitudinal and transversal (with respect to the neutrino
momentum) components of the vector $\mathbf{G}$,
$\mathbf{n}_{\perp}$ is the unite orthogonal to $\mathbf{n}$ and
laying in one plane with $\mathbf{n}$ and $\mathbf{L}$ vector. In
order to get the probability of the neutrino spin oscillations in
the gravitational field we introduce the unit vector $\mathbf{k}$

\[
  \mathbf{k} = \frac{\mathbf{G}}{G},
\]
and rewrite eq.\eqref{eq} in the form
\begin{equation}\label{eqnew}
  \frac{d\bm{\zeta}_\nu}{dt}=
  \frac{\alpha}{r^3}
  [\bm{\zeta}_\nu\times\mathbf{k}],
\end{equation}
where with the use of (\ref{G}), (\ref{Gpar}), and (\ref{Gper}) we
have
\begin{equation*}
  \alpha=\frac{G_N L}{\gamma}
  \sqrt{4\gamma^2\cos^2\theta+\sin^2\theta}.
\end{equation*}
If we expand the neutrino spin vector ${\bm \zeta}_{\nu}$ over the
basis determined by the vector $\bf G$, so that
\[
  \bm{\zeta}_\nu=\bm{\zeta}^{\perp}_\nu+
  \bm{\zeta}^{\parallel}_\nu,
\]
where
   $(\bm{\zeta}^{\perp}_\nu
\mathbf{k})=0$ and $(\bm{\zeta}^{\parallel}_\nu
\bm{\zeta}^{\perp}_\nu)=0$, then the solution of eq.\eqref{eqnew}
in terms of $ \bm{\zeta}^{\parallel}_\nu$,\ $\bm{\zeta}_{x}$ and
$\bm{\zeta}_{y}$ ($\bm
\zeta^{\perp}_\nu=\sqrt{\zeta_x^2+\zeta_y^2}$), is given by
\begin{align}\label{solution}
  \zeta_x &
  =-\sqrt{1-\zeta_0^2}\sin
  \left(
  \frac{\alpha}{2\beta r^2}
  \right),
  \notag
  \\
   \zeta_y &
  =\sqrt{1-\zeta_0^2}\cos
  \left(
  \frac{\alpha}{2\beta r^2}
  \right),
   \\
  \zeta^\parallel_\nu &
  =\zeta_0.
  \notag
\end{align}
Here $\zeta_0$ is a constant determined by the initial conditions.
Finally, for the probability of the neutrino oscillations in the
gravitational field, described by the vector $\bf G$, we get
\begin{equation}\label{probab}
  P_{\nu_L\to\nu_R}(t)=
  \frac{1}{2}[1+(\bm{\zeta}_\nu \mathbf{n})],
\end{equation}
were $\bm{\zeta}_{\nu}$ is the solution of eq.\eqref{eqnew} given
by \eqref{solution}.

\section{Spin light of neutrino in gravitational fields}

From our previous studies \cite{LobStuPLB03} of the $SL\nu$ we
know that this radiation is emitted if the neutrino magnetic
moment is processing, no matter what is the cause of this
procession. Therefore, using the results for the $SL\nu$ in matter
and electromagnetic fields and the analogy between neutrino spin
evolution eqs.~(\ref{eq}) and (\ref{S}) we get for the total power
of the {\it spin light of neutrino in the gravitational field}
\begin{equation}
  I_\mathrm{gr}=\frac{16}{3}\mu^2
  \left[
  4(\mathbf{G}^2)^2+\dot{\mathbf{G}}^2
  \right],
\label{IG}
\end{equation}
where $\mu$ is the neutrino magnetic moment.

Let us now consider the new phenomenon of the $SL\nu$ in the case
of neutrinos moving in the gravitational fields of a rotating
neutron star along its radial direction. From eqs. (\ref{G}),
(\ref{Gpar}), and (\ref{Gper}) it is easy to see that for the
ultra relativistic neutrinos ( for the neutrinos from neutron
stars the Lorentz factor could be of the order of $\gamma \sim
10^9$) the total power of the $SL\nu$ is maximal when neutrinos
move along or opposite to the angular momentum $\mathbf{L}$:
\begin{equation}
\displaystyle   I_\mathrm{gr}=48\mu^2\gamma^4\frac{G_N^2L^2}{r^8}
       \left[
       {\frac{1}{1-\frac{1}{\gamma^2}}}+\frac{G_N^2L^2}{9r^4}
       \right].
\label{IGmax}
\end{equation}
For a neutron star with the radius $\sim10 \ \text{km}$ we get

\begin{equation}
  I_\mathrm{gr}=3\mu^2\gamma^4
  \left[
  1+\frac{1}{144}\left(\frac{1\thinspace\text{km}}{r}\right)^4
  \right]
  \frac{1}{r^8}.
\end{equation}
It follows that the total radiation power of the $SL\nu$ in the
gravitational field is proportional to $\sim \gamma ^4$ and the
main contribution is given by the second term in (\ref{IGmax})
which originates from the derivative term in the general
expression (\ref{IG}). It is also possible to show that there is a
strong beaming effect and that the radiation is confined within a
small cone in the direction of the neutrino propagation given by
the angle $\delta\theta\sim\gamma^{-1}$. Thus, we predict that the
angular distribution of the $SL\nu$ in the gravitational field of
a rotating neutron star is not isotropic even if neutrinos are
moving symmetrically in all radial directions from the neutron
star. In this case the  $SL\nu$ in the gravitational field is
radiated more effectively by neutrinos moving along or against the
axis of the neutron star rotation.

It is also possible to consider the combining effect of the
electromagnetic, weak and gravitational fields on neutrino spin
procession. From (\ref{eq}) and (\ref{S}) we get for the neutrino
spin evolution in this case
\begin{equation}
   \frac{d\bm{\zeta}_\nu}{dt}=
   \frac{2}{\gamma}[\bm{\zeta}_\nu\times\left(
   \mu \mathbf{B_0}+\mu \mathbf{M_0}+\mathbf{G}\right)],
\end{equation}
where  $\mathbf{B_0}$, $\mathbf{M_0}$  are given by
eqs.(\ref{B_0}) and (\ref{M_0}) respectively. The gravity effect
enters through the vector $\mathbf{G}$, which in the considered
case of a rotating neutron star, is given by eq.(\ref{G}).

 It is interesting to compare the contributions to the
$SL\nu$ from the neutrino interaction with the background matter
and with the background gravitational field. For illustration, let
us investigate the two cases: i) the $SL\nu$ from a rotating black
hole produced by neutrinos moving in accretion disk, and ii) the
$SL\nu$ from a quasar produced by neutrinos moving along the
relativistic jet. We take matter consisting of hydrogen. In the
first case we assume that matter in the accretion disk is of
constant density and slowly moving along the radial direction. In
the second case we consider matter of the jet moving with
relativistic speed. We omit here the electromagnetic interaction
but account for  the neutrino couplings with matter and
gravitational field, aiming to consider also the interplay between
them. This modelling of the background environments, although is
rather simplified, could reveal the main features of the effects
under consideration.

We start with the general expression \cite{DvoStuJHEP02} for the
neutrino spin evolution Hamiltonian  and extract for the further
consideration the terms accounting for the neutrino interactions
with the gravitational field and matter:
\begin{equation}\label{H}
  \mathcal{H}=
  g_a(\bm{\Sigma}\mathbf{A})+
  \left(\gamma_{5}f^0(x)+(\bm {\Sigma}\,\mathbf {f}(x))\right),
\end{equation}
where $g_a\mathbf{A}$ is given by (\ref{A}). The matter term
contribution in (\ref{H}) is determined in the general form as
\begin{equation}
  \label{f}
  f^{\mu}=\frac{G_F}{\sqrt2}\sum\limits_{f=e,p,n}
  \left(j^{\mu}_{f}q^{(1)}_{f}+\lambda^{\mu}_{f}q^{(2)}_{f}\right),
\end{equation}
где
\begin{align}
  \label{q1q2}
  q^{(1)}_{f}&=
  (I_{3L}^{(f)}-2Q^{(f)}\sin^{2}\theta_{W}+\delta_{ef}),
  \quad
  q^{(2)}_{f}=
  -(I_{3L}^{(f)}+\delta_{ef}),
  \\
  \delta_{ef}&=
   \begin{cases}
     1, & \text{$f=e$,}
     \\
     0, & \text{$f=n,$ $p$,}
   \end{cases}
  \notag
\end{align}
however in the considered case of matter composed of hydrogen
there is no contributions from neutrons $n$. In this equations,
$I_{3L}^{(f)}$ denotes the third component of weak isospin for the
fermion $f$, and $Q^{(f)}$ is the electric charge of the fermion.
The matter current and polarization are \cite{LobStuPLB01},
respectively,
\begin{align}\label{j}
  j^{\mu}&=(n_f, n_f \mathbf{v}_f),
  \\
  \label{lambda_e}
  \lambda^\mu &=
  \left(
  n_f \bm{\zeta}_f \mathbf{v}_f ,
  n_f \bm{\zeta}_f \sqrt{1-v^2_f}+
  \frac{n_f\mathbf{v}_f
  \left(
  \bm{\zeta}_f \mathbf{v}_f
  \right)}{1+\sqrt{1-v^2_f}}
  \right),
  f=e, p.
\end{align}

Finally, for the neutrino spin evolution equation in the
considered case we have
\begin{equation}\label{eq_m_g}
  \frac{d\bm{\zeta}_\nu}{dt}=\frac{2}{\gamma}\left[
  \bm{\zeta}\times(\mu \mathbf{M}_0+\mathbf{G})\right],
\end{equation}
where the matter term according to eqs.(\ref{H}), (\ref{f}), and
(\ref{q1q2}) is
\begin{equation}
  {\mathbf{M}_0}=
  \gamma\bm{\beta}
  \left(
  f^0-
  \frac{\gamma}{1+\gamma}
  (\mathbf{f}\bm{\beta})
  \right)-
  \mathbf{f},
\end{equation}
and the gravitational field effect is described by the vector
$\mathbf{G}$ which is given by (\ref{G}). Thus, for the total
radiation power of the $SL\nu$ in matter and the gravitational
filed we find,
\begin{equation}
  I_\mathrm{gr,matt}=\frac{16}{3}\mu^2
  \left[
  4\big((\mu\mathbf{M}_0+\mathbf{G})^2\big)^2+\big(
  \mu \dot{\mathbf{M}}_0+\dot{\mathbf{G}}\big)^2
  \right].
\end{equation}
It is evident that the total radiation power,
$I_\mathrm{gr,matt}$, is composed of the three contributions,
\begin{equation}
  I_\mathrm{gr,matt}=I_\mathrm{gr}+I_\mathrm{matt}+I_\mathrm{gr+matt},
\end{equation}
where $I_\mathrm{gr}$ is the radiation power due to the neutrino
magnetic moment interaction with the gravitational field,
$I_\mathrm{matt}$ is the radiation power due to the neutrino weak
interaction with the background matter, and $I_\mathrm{gr+matt}$
stands for the interference effect of the gravitational and weak
interactions. In an analogy with the neutrino spin flip in matter
(see \cite{LobStuPLB03}) we predict the similar effect for
neutrino spin in gravitational fields: the initially unpolarized
neutrino beam (equal mixture of active left-handed and sterile
right-handed neutrinos) can be converted to the totally polarized
beam composed of only $\nu_{R}$.

Let us turn to the discussion on the $SL\nu$ in the two particular
cases mentioned above:

i)~It is obvious that in this case at small distances $r$ from the
center of the black hole the  contribution to the $SL\nu$ coming
from the gravitational interaction dominates over that from the
interaction with matter. As neutrino propagates away from the
center, the gravity contribution fades and starting from a certain
distance from the black hole center the matter interaction
contribution to the $SL\nu$ becomes predominant. The total
radiation power, $I_\mathrm{gr,matt}=I(r)$,  versus the distance
$r$ is shown in Fig.~\ref{fig}. The green dash-dotted line
corresponds to the matter contribution, while the blue dashed line
stands for the gravitational field contribution to the $SL\nu$.
For the neutrino gamma factor, neutrino magnetic moment and matter
density we take, respectively, the following values of
$\gamma=10^{12}$, $\mu=10^{-10}\mu_0$ and
$n=10^{24}\thinspace\text{cm}^{-3}$, where $\mu_0$ is the Bohr
magneton. It is supposed that the gravitational field is produced
by the rotating object with the solar mass $M=M_\odot$, and the
angular momentum is chosen to be equal to the maximal allowed
value $L=r_0^2/(4G_N)$ ( see, for instance,
\cite{Lan_Lif_Nauka88})  where $r_0$ is the Schwarzschild radius.
\begin{figure}
  \centering
  \includegraphics[scale=.6]{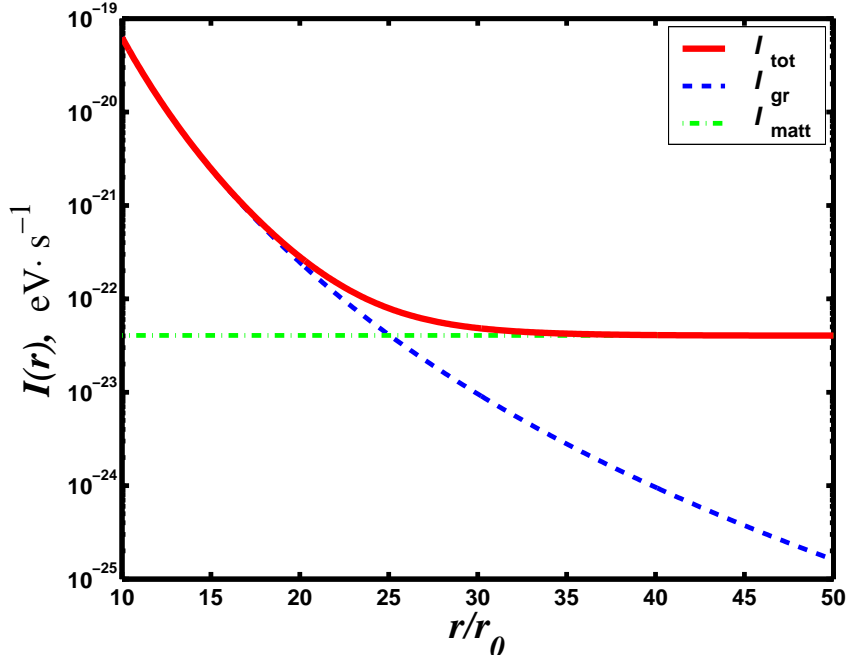}
  \caption{The total radiation power of the spin light of neutrino,
  $I(r)=I_\mathrm{gr}+
  \ I_\mathrm{matt}+\ I_\mathrm{gr+matt}$, versus radial distance $r$ is shown
  by the red solid line. The blue dashed and green dash-dotted
  lines correspond, respectively, to the gravitational, $I_\mathrm{gr}$, and
  matter, $I_\mathrm{matt}$, contributions without account
  for the interference term $I_\mathrm{gr+matt}$.}

  \label{fig}
\end{figure}

ii)~In this case, following eq.(\ref{eq_m_g}) the total radiation
power is
\begin{equation}
  I_\mathrm{gr,matt}=
  \frac{64}{3}\mu^2\gamma^4
  \left[
  \left(
  {\frac{G_N\mathbf{L}}{r^3}}-{\frac{G_F}{\sqrt2}}\sum_{f=e,p}{n_{f}
  (\mathbf{v}_{f}-\bm{\beta})}
  \right)^4
  +\frac{9}{4}{\frac{G^2_N\beta^2L^2}{r^8}}
  \right].
\end{equation}
If matter is moving with relativistic speed along the direction of
the neutrino propagation then the matter contribution to
$I_\mathrm{gr,matt}$ is washed out because of the presence of the
term $(\mathbf{v}_{f}-\bm{\beta})\sim 0$ . This happens due to the
particular properties of the neutrino oscillations in moving
matter (see
\cite{LikStu95,LobStuPLB01,GriLobStuPLB02,DvoStuJHEP02, StuYF04}).
Thus, we conclude that in the case of the neutrinos moving in the
relativistic jet of a quasar the $SL\nu$ could be produced by the
gravitational field, whereas there is no important contribution
from interaction with matter.

\section{$\bm S\bm L\bm \nu$ photon energy in gravitational fields}

Now we discuss the energy range of the $SL\nu$ photons emitted in
a background where only the presence of the gravitational field of
a rotating object generates the neutrino spin procession (it is
supposed that neither matter nor electromagnetic fields of the
background gives important contribution to the neutrino spin
evolution). In order to get an estimation for the characteristic
scale of the energy of the emitted photons we suppose that the
variation of the vector $\mathbf{G}$ (which describes the
influence of the gravitational field) on the neutrino travelled
distance $\Delta r$ is much less then the frequency of the
neutrino spin precession,
\begin{equation}
  \frac{\Delta G}{G}\ll\omega\Delta r.
\end{equation}
If we consider the neutrino propagating along the rotation axis,
$\theta=0$, and also take into account followed from
eqs.\eqref{Gpar} and \eqref{Gper} relation $G_\parallel\gg
G_\perp$ then the emitted photon energy in the rest frame of the
neutrino is
\begin{equation}\label{omega0}
  \omega_0\sim\frac{G_{N}L}{r^3}\gamma.
\end{equation}
To calculate the photon energy in the laboratory reference frame
we should use the well known formula
\begin{equation}\label{omegarecalc}
  \omega=\omega_0\frac{\sqrt{1-\beta^2}}{1-\beta\cos\vartheta},
\end{equation}
where $\vartheta$ is the angle between the neutrino speed and the
direction towards the observer. Due to the fact that the $SL\nu$
is strongly beamed within a small cone there is a reason to
consider the case of $\vartheta\approx 0$. Therefore, the emitted
photon energy in the laboratory frame is
\begin{equation}\label{omega}
  \omega\sim\omega_0 \gamma \sim
  \frac{G_{N}L}{r^3}\gamma^2.
\end{equation}
For the set of parameters that we have just considered in the case
of a rotating black hole, eq.(\ref{omega}) reads
\begin{equation}\label{omegavalue}
  \omega\sim 10^{-11}\times \gamma^2
  \left(
  \frac{r_0}{r}
  \right)^3
  \thinspace\text{eV}.
\end{equation}
For $\gamma \sim 10^{12}$ and $r \sim 10 \ r_0$ we obtain
$\omega\sim 10\thinspace\text{GeV}$. Note that for the mass of
neutrino $m_\nu\sim 1\thinspace\text{eV}$ we also obtain
$(\omega/E_\nu)\sim 10^{-2}\ll 1$, i.e. the quasiclassical
approach to the neutrino spin evolution is valid in this case.
These properties of $SL\nu$ enable us to predict that this
radiation should be important in different astrophysical
environments (quasars, gamma-ray bursts etc) and in dense plasma
of the early Universe.

\section{Conclusion}

We also should like to point out that although the predicted above
and in \cite{LobStuPLB03} the {\it spin light of neutrino} in
matter, electromagnetic and gravitational fields is supposed to be
radiated in the process
\begin{equation}\label{process}
  \nu_1 \rightarrow \nu_2 + \gamma,
\end{equation}
without change of the neutrino flavour state (the neutrinos in the
initial and final states, $\nu_1$ and $\nu_2$, are of the same
flavour), it is possible to generalize, with a minor modification,
the corresponding equations for the case when the $SL\nu$ is
radiated due to the neutrino transition magnetic moment
interaction with the background fields. The latter corresponds to
models of the Dirac neutrinos with non-diagonal magnetic moments
or the Majorana neutrinos which could also have transitional
(magnetic) moments.

In conclusion, we consider the neutrino spin evolution in presence
of gravitational fields and derive for the first time the
corresponding neutrino oscillations probability. We also predict
the possibility of a new mechanism of the spin light of neutrino
($SL\nu$) in a gravitational field which, as we expect, could have
consequences in astrophysical and cosmological settings. As
examples, we analyzed the $SL\nu$ in particular cases of neutrino
moving in the vicinity of a rotating neutron star, black hole and
the relativistic jet of a quasar.

\end{document}